# Bose-Einstein condensation: what, how and beyond

Vishwamittar

Department of Physics, Panjab University, Chandigarh – 160014, India.

Email address: vm@pu.ac.in

**Abstract**

The piling up of a macroscopic fraction of noninteracting bosons in the lowest energy state of a system at very low temperatures is known as Bose-Einstein condensation. It took nearly 70 years to observe the condensate after their theoretical prediction. A brief history of the relevant developments, essentials of the basic theory, physics of the steps involved in producing the condensate in a gas of alkali atoms together with the pertinent theory, and some important features of the research work carried out in the last about 25 years have been dealt with. An effort has been made to present the material in a manner that it can be easily followed by undergraduate students as well as non-specialists and may even be used for classroom teaching.

## 1. INTRODUCTION

Bose-Einstein condensation (BEC) is the fascinating phenomenon of cumulation of a very large fraction of identical boson particles (the entities having integer spin, which is actually an integer multiple of $\hbar$, the Planck's constant of action divided by $2\pi$, and described by symmetric wavefunction) in the lowest energy or the ground state in a system under appropriate characteristic conditions of temperature, number density, etc. It represents a phase transition to a state of matter in which a good number of constituents of the system suddenly coalesce into a single coherent quantum mechanical entity that can be described by a wavefunction on nearly macroscopic scale and the presence of interparticle interactions is not a prerequisite. It is completely different from the conventional gas to liquid condensation, like steam to water, where the interparticle attractions play an important role. As such, BEC is a manifestation of the quantum effects arising from the indistinguishability of the particles with integer spin leading to their distribution being governed by a specific quantum-statistics. The condensate created in the process is an extremely-low-density condensed matter system.

The history of BEC is intimately connected with the early developments in quantum mechanics. In the year 1900, Planck explained the then puzzling results for the blackbody radiation spectra by introducing the revolutionary idea that a blackbody emits radiation in discrete energy packets or quanta, which were later identified as particles having zero rest mass and, in 1926, were given the name photons. Einstein (1905) asserted that these radiations are not only emitted in quanta but also interact as packets and thereby explained the empirical laws of photoelectric effect.

While teaching the above-mentioned topics, Bose (1924) obtained the famous Planck's blackbody radiation formula using a purely statistics-based treatment for the radiation quanta whose total number is not conserved because of continuously repeated emission from and absorption by the walls of the blackbody. It is important to note that though his approach was indigenous and completely different from the then available Maxwell–Boltzmann statistics (derived assuming the particles to be distinguishable), yet his research paper was rejected by a British journal. Then he sent his work to Einstein who was so impressed with the concept that he translated the manuscript from English to German and got it published, in Bose's name, in a journal there. Not only that, Einstein (1924-25) immediately extended the ideas to the case of ideal particles with nonzero rest mass, whose total number is conserved, and brought out two papers in a short period. Thus, a new form of statistics was born and, later it was found that all the particles with integer spin follow its distribution law. This formalism came to be known as Bose–Einstein statistics and the particles obeying this are referred to as bosons.

Einstein showed that in a gas of noninteracting particles, at sufficiently low temperature and having proper number density, a significant fraction of these will occupy the lowest energy single–particle state as if these have undergone a transition from the gaseous form to the condensed form. However, to begin with, this conjecture, now known as Bose–Einstein condensation, was not taken seriously. It was in 1938 that London suggested that the then recently observed superfluidity in liquid $^4$He below 2.18 K, could be understood in terms of BEC and this was followed by some theoretical works in this direction. Later in 1957, superconductivity in some metals at very low temperatures was also attributed to BEC of Cooper pairs formed by two opposite spin electrons. However, in spite of some success in explaining different experiments and research works on the connection between BEC and these two phenomena, both these examples suffered from a big drawback that BEC was predicted to occur in an ideal or noninteracting gas, while the existence of both the liquid helium and the Cooper pairs is not possible without an interaction. Consequently, the urge to search for its realization in pure form, much closer to the original idea of BEC, persisted.

The first serious effort in the direction of obtaining BEC in an ideal gas began in 1978, when Greytak and Kleppner with their co-workers started looking for this phenomenon in spin polarized atomic hydrogen because it had been shown by Hecht (1959) and later by Stwalley and Nosanow (1976) that it would remain a gas down to 0 K without forming a liquid or solid. In the meantime, sophisticated techniques to attain extremely low temperatures and for manipulating atoms were being developed and continuously improved in many laboratories, and, by 1989, it had been realized that alkali atomic gases would be more suited for observing BEC. These dedicated efforts bore fruits in the middle of 1995 when unambiguous experimental evidence for existence of BE condensate (a *super atom*) was reported from three laboratories in quick succession. Cornell, Wieman and their co-workers created this in a gas of $^{87}$Rb atoms (cooled to 170 nK and below) in June 1995; Hulet and his colleagues formed BEC in spin polarized $^7$Li atoms in July 1995; and Ketterle and his team obtained the condensate in a gas of $^{23}$Na atoms in September 1995. The importance of these milestone experiments producing 'coldest atoms in the Universe' was immediately highlighted by Science magazine (published by American Association for the Advancement of Science) by announcing the condensate as 'molecule of the year' and depicting this as a platoon of soldiers marching in complete unison, on the cover of the Dec 22, 1995 issue. Furthermore, Cornell, Ketterle and Wieman were honoured with the 2001 Nobel Prize in physics 'for the achievement of Bose – Einstein condensation in dilute gases of alkali atoms, and for early fundamental studies of the properties of the condensates' [1-9].

This spectacular observation in 1995 aroused immense interest in BEC and since then, a wide variety of experiments pertaining to its different facets have been carried out in numerous laboratories around the world, and condensates have been produced in a good number of atomic species, including spin polarized atomic hydrogen, and, also, in some molecules, quasiparticles and photons. Besides, various theoretical aspects too are being investigated with full zeal. The research pursuits in this field can be, in general, said to belong to atomic physics, condensed matter physics, quantum optics and statistical physics depending on the emphasis of the work.

In this article, we briefly describe the essentials of theoretical aspects needed to understand the phenomenon and the main features of the experimental steps involved in obtaining the condensate in a dilute gas of ultra-cold alkali atoms. Also included are some recent important results and possible applications of BEC.

## 2. GENERAL BASICS

Suppose we have a gas of noninteracting or ideal identical bosons of mass m, in equilibrium in a container of volume $V$ maintained at temperature $T$. We identify the single–particle quantum energy states of the system by $\varepsilon_j$; $j = 0, 1, 2, \ldots$. It is assumed that the energy eigenstates are nondegenerate and that all the particles are in the same spin state so that contribution of this aspect can be suppressed. The bosons do not obey Pauli's exclusion principle so that there is no restriction on the number of particles that can occupy the same eigenenergy state. The average number of bosons occupying the energy state $\epsilon_j$ is given by the distribution law [3, 9]

$$<n_j> = \frac{1}{e^{\beta(\varepsilon_j - \mu)} - 1}. \tag{1}$$

$<n_j>$ is called occupation number and $\beta = \frac{1}{kT}$, with k as Boltzmann constant, is the temperature parameter. It is conventional to introduce a parameter $\zeta = e^{\beta\mu}$, known as absolute activity or fugacity of the system, so that Eq. (1) becomes

$$<n_j> = \frac{1}{\zeta^{-1} e^{\beta\varepsilon_j} - 1}. \tag{2}$$

Here, $\mu$ is the chemical potential and is implicitly defined by the condition

$$\sum_j <n_j> = N, \tag{3}$$

the total number of particles in the system and explicitly depends on T besides N. Physically speaking, it is change in internal energy of the system brought about by addition of a particle.

In view of the fact that the occupation number cannot be negative, we must ensure that $e^{\beta(\varepsilon_j - \mu)} \geq 1$ for all temperatures and energy states of the system. This demands that $\mu$ should be such that the condition $\varepsilon_j \geq \mu$ is always satisfied. Obviously, this will be so if it holds for the lowest energy state, that is, $\varepsilon_0 \geq \mu$, implying that $\mu$ can never exceed $\varepsilon_0$. Taking the ground state energy as 0, this implies that for a boson gas, the chemical potential is necessarily negative and at most it can have value 0, whatever the temperature T. Also, from the fact that $\beta\mu \leq 0$, we have $\zeta = e^{\beta\mu} \leq 1$; i.e., $\zeta$ is a positive fraction. It may, however, be mentioned that $\mu = 0$ for the bosons for which the particle number is not conserved.

The exact expression for the temperature dependence of $\mu$ shows that for a boson gas at sufficiently low temperatures, $\mu \approx -\frac{kT}{N} = -1.38 \times 10^{-23}$(T/N), implying that $\mu$ has very small magnitude. Thus, $\beta\mu \approx -\frac{1}{N}$, so that $\zeta \approx e^{-\frac{1}{N}} \approx 1 - \frac{1}{N}$ and $\zeta^{-1} \approx e^{\frac{1}{N}} \approx 1 + \frac{1}{N}$, for large $N$, making $\zeta$ and $\zeta^{-1}$ to be close to unity for actual systems at low temperatures. This inference together with Eq. (2) shows that the occupation number for the lowest energy state $\varepsilon_0 = 0$, can be very large, particularly at low temperatures. Therefore, while dealing with the expression for $N$, namely Eq. (3,) we separate the ground state energy term from the other terms accounting for the occupancy of the excited states and write

$$N = \frac{1}{\zeta^{-1}-1} + \sum_{j,j\neq 0} \frac{1}{\zeta^{-1}e^{\beta\varepsilon_j}-1} = N_0 + N_{ex}. \qquad (4)$$

Here, $N_0 = \frac{1}{\zeta^{-1}-1}$ gives the number of bosons in the $\varepsilon_0 = 0$ single–particle energy state, out of the total number, $N$, and can be extremely large because the denominator in the expression is of the order of $\frac{1}{N}$. A detailed theoretical treatment shows that in the thermodynamic limit $N \to \infty, V \to \infty$ such that their ratio is always finite, the maximum number of bosons that can be accommodated in the excited energy states is given by

$$(N_{ex})_{max} = 2.612 \frac{V}{\lambda^3}, \qquad (5)$$

where

$$\lambda = \frac{h}{\sqrt{2\pi m k T}} \qquad (6)$$

and is referred to as the mean thermal wavelength of the particles. Note that a particle with thermal kinetic energy $3kT/2$, has linear momentum $p = \sqrt{\frac{(2m)3kT}{2}}$ so that

$$\lambda = \sqrt{\frac{3}{2\pi}} \left(\frac{h}{p}\right) = \sqrt{\frac{3}{2\pi}} \Lambda_{dB}. \qquad (7)$$

Here, $\Lambda_{dB}$ is thermal de Broglie wavelength associated with the particles and provides a measure of the length over which the particle wavefunction or wave packet extends. It may be mentioned that many physicists refer to $\lambda$ itself as de Broglie wavelength. Substituting Eq. (7) into Eq. (5), we get

$$(N_{ex})_{max} = 7.917 \frac{V}{\Lambda_{dB}^3}. \qquad (8)$$

As long as the total number of bosons $N$ in the sample is less than $(N_{ex})_{max}$, all the particles are distributed in different energy states. However, if $N$ exceeds $(N_{ex})_{max}$, then the excited energy states accommodate $(N_{ex})_{max}$ particles and the remaining $N - (N_{ex})_{max}$ pile up together into the $\varepsilon_0 = 0$ state. Thus,

$$N_0 = \frac{\zeta}{1-\zeta} = N - 2.612 \frac{V}{\lambda^3} = N - 7.917 \frac{V}{\Lambda_{dB}^3}. \qquad (9)$$

It is worth pointing out that $(N_{ex})_{max}$ is proportional to $T^{3/2}$ so that for a specific system, this number decreases and $N_0$ increases as temperature is lowered. Moreover, $N_0$ is an anomalously large number. This accumulation of enormously large number of bosons in the single–particle ground state is *Bose–Einstein condensation*. This transition differs from the condensation of vapours into the liquid state on two accounts. Firstly, the derivation of the last result is for an ideal quantum-mechanical gas obeying BE statistics, completely free from any interactions and, thus, this is a purely quantum effect. Secondly, this involves collection of a large fraction of bosons in the zero–energy state, and, hence, separation of the zero–momentum particles from the remaining particles, meaning that the condensation is in the momentum space rather than the usual coordinate space.

From Eq. (9), we find that a gas will undergo BEC only if $N_0 > 0$. For a given system, in terms of temperature, this requires that

$$T < T_c = \frac{h^2}{2\pi mk} \left(\frac{N}{2.612\, V}\right)^{2/3}. \tag{10}$$

$T_c$ is the critical temperature for transition from the normal state to BEC state, and this will have higher value if the mass of the bosons is small and the particle number density ($n = N/V$) is large. Also, $T_c$ is related to the mean thermal wavelength of the particles as

$$T_c = \lambda^2 T \left(\frac{N}{2.612\, V}\right)^{2/3}. \tag{11}$$

Combining Eqs. (9) and (11), we get

$$\frac{N_0}{N} = 1 - \left(\frac{T}{T_c}\right)^{3/2}. \tag{12}$$

This shows that the fraction of the bosons in the condensate will be close to unity if the temperature $T$, to which the system is cooled, is significantly low and the critical temperature is high. Thus, at temperatures lower than $T_c$, the system consists of two parts: BE condensate having $N_0$ particles and $N - N_0$ normal or above-condensate particles. For the condensate, chemical potential is zero because of the natural tendency of the bosons to populate the lowest energy state. In fact, as a function of temperature $\mu$ increases (retaining its negative value) with decrease in temperature till it becomes 0 at $T_c$ and continues to be so at lower temperatures.

As mentioned in the introduction, the liquid He I to He II transition observed at 2.18 K was thought to be a manifestation of BEC. For this, substituting $m = 6.65 \times 10^{-24}$g, $V = 27.6$ cm$^3$/mole, and $N = 6.02 \times 10^{23}$ mole$^{-1}$ into Eq. (10), we get $T_c = 3.14$ K. Its difference of about 44% from the experimental value, was attributed to the presence of interactions and, later, it turned out that inclusion of interactions did improve agreement for some of the experimental results. However, many observed phenomena could not be accounted for and it was found from the neutron scattering experiments that only about 10% of the atoms undergo transition, which is at great variance with the predictions.

Eq. (9) shows that the condition for obtaining BEC can also be expressed as

$$N > 7.917 \frac{V}{\Lambda_{dB}^3} \quad \text{or} \quad \Lambda_{dB} > 2\left(\frac{V}{N}\right)^{1/3}; \tag{13}$$

that is, number density should be above a critical value $\frac{7.9}{\Lambda_{dB}^3}$. Note that de Broglie wavelength $\Lambda_{dB}$ is the average size of the wave packet associated with the particles and $\left(\frac{V}{N}\right)^{1/3}$ is the mean separation of the particles in the box. Accordingly, the above equation implies that BEC can occur only if the number density, $n = N/V$, of the bosons in the gas is so high that the interparticle separation is less than half their de Broglie wavelength at the prevalent temperature. Of course, the particle number density should still be such that the collection is effectively a noninteracting gas. But the size of the wave packet being more than the distance between the particles means that these overlap each other. Thus, BEC is a consequence of the overlap of the wave packets or the wavefunctions of the bosons. That is why, we say that BEC is a quantum-mechanical effect. Furthermore, in the condensate, $N_0$ particles are in the same state and overlap of their wavefunctions means that all of these can be described by a single

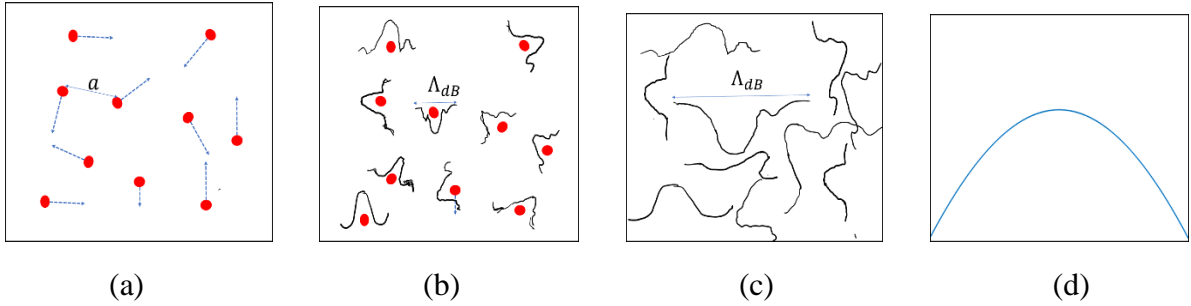

(a)          (b)          (c)          (d)

**Fig. 1.** Boson gas in a box at different temperatures: (a) Particle nature at a high temperature; (b) Wave packet representation at low temperatures; (c) Wave packets start overlapping indicating onset of BE condensation at critical temperature; (d) Condensate is described by a single wave packet at very low temperatures.

wavefunction – a sort of long–range order in the noninteracting gas. This justifies the condensate being referred to as a super atom.

It may be mentioned that $n = N/V$ is the number density in ordinary or coordinate space and $\Lambda_{dB}$, being defined in terms of momentum, makes this a momentum space quantity. The product $n\Lambda_{dB}^3$ is said to give phase space density so that Eq. (13) can be interpreted as a requirement that phase space density should be more than 8 for the occurrence of BEC.

In order to elaborate the above statements, prompted by a slide in Ketterle's C. N. Yang symposium at Stony Brook (May 21, 1999), we consider the following pictorial representation of the boson particles in a box at different temperatures; Fig. 1. At reasonably high temperatures, the particles behave like classical distinguishable entities or marbles, moving randomly with different velocities governed by Maxwell-Boltzmann distribution; Fig. 1(a). At these temperatures, velocities and hence momenta of the particles are so large that associated wave packets have $\Lambda_{dB}$ ($\approx 10^{-2}$ nm) less than the particle dimensions and much less than the average spacing, $a$, between the particles ($\approx 10^2$ nm). Accordingly, the wave packets are substantially separated and play no role in describing the trajectories of the particles. As the temperature is lowered, velocities decrease rendering $\Lambda_{dB}$ to be larger than the particle size as projected in Fig. 1(b). Thus, quantum aspect becomes necessary, and the particles are described by appropriate wavefunctions. At still lower temperatures, the randomness and velocities of the particles decrease significantly making momentum very small and $\Lambda_{dB}$ to be quite large. This results in overlap of the wave packets associated with neighbouring particles, making quantum effects to be important. This situation is shown in Fig. 1(c). The temperature at which the overlap starts, i. e., the de Broglie wavelength is of the order of interparticle separation, is the critical temperature for formation of BEC. At extremely low temperatures, the $\Lambda_{dB}$ values become comparable with the linear dimensions of the box and overlap of the wave packets is so strong that we cannot distinguish the particles, and these can be described by the same wavefunction as if this were a single quantum entity. In other words, the wave packet becomes macroscopic, and we have the condensate state of completely coherent particles; Fig. 1(d).

## 3. CHOICE OF BOSON GAS FOR BEC

The main lesson learnt from the discussion in Section 2 is that to observe BEC, the collection of bosons must have number density large enough to make their spacing much smaller than their de Broglie wavelength and still be free from any interactions which could overshadow the quantum effects. This is quite a stringent condition and satisfying this, posed a big challenge to physicists as this required working at extremely low temperatures with bosons in the gaseous state and, also, to avoid adsorption of the particles at the surfaces of the container. This not only involved development of numerous indigenous precise techniques but also perfecting these. That is why there was a time span of about 70 years between the prediction and experimental realization of BEC.

We know that all the material objects consist of atoms, which themselves are made of fermions – electrons, protons, and neutrons; all having half-integer spin. An atom has characteristic atomic number Z and a mass number A. For the bosons to be considered for BEC, these must be devoid of an electrical charge. Now, an atom has Z electrons, Z protons and A – Z neutrons. Since total spin of a collection of an even number of half-spin particles will be an integer, an atom behaves as a boson or a fermion depending on whether A – Z is an even or an odd number. Therefore, for a neutral atom to be a boson, both A and Z should be either even or odd numbers so that A – Z, the number of neutrons, is even.

The first obvious choice for creating BEC was hydrogen ($^1$H) atom, for which both Z and A are unity. Besides, it had been argued that when a collection of $^1$H atoms is subjected to a strong magnetic field, spins of most of the atoms get aligned (the spin polarized atomic hydrogen) and that their attractive interaction is so weak that this collection would continue to be a gas at all temperatures. Though the experimental work on this boson gas started long back, it took nearly 20 years to yield the expected results.

In the meantime, the emphasis shifted to neutral alkali atoms because these have simple ground state electronic configuration, and their collection remains a gas down to very low temperatures if their number density does not exceed $10^{15}$ atoms cm$^{-3}$. Moreover, cooling these was relatively easy as compared to the spin polarized hydrogen. Since alkali atoms have odd value of Z, isotopes having odd A were the natural choice. In fact, the early successful demonstrations for existence of BEC were made nearly 27 years back in the atomic gases of $^{87}$Rb, spin polarized $^7$Li and $^{23}$Na. Note that for $^{87}$Rb, if we take $\frac{N}{V}$ = 2.5 x $10^{12}$ cm$^{-3}$, then Eq. (1) gives $T_c$ = 34 nK.

## 4. EXPERIMENTAL ASPECTS

Having dwelt upon the basic theory of BEC and reasons for the preference of dilute gas boson alkali atoms for its experimental realization, we briefly talk about the main steps involved in obtaining the condensate in these. To avoid making description very technical, we just discuss the essential physics behind relevant components of the experimental set up and the techniques employed. It may be mentioned that the approach followed in getting BEC in spin polarized hydrogen gas is quite different and is not dealt with here.

It must be noted that the atomic gas being used for obtaining BEC cannot be stored in a conventional container because it is not that easy to make it so cold and even if this is done, the atoms would stick to the walls. This problem is solved by using a small ultrahigh vacuum glass cell with arrangement for creating different facilities for trapping as well as successive cooling at appropriate stage. In fact, the same system is used to serve both the purposes, making sure that the atoms do not come in contact with walls of the cell, which itself is maintained at room temperature [1-8].

## 4.1. Oven

It is maintained at such a temperature that the alkali metal evaporates and has desired number density n. For sodium, the temperature is about 600 K and $n \approx 10^{14}$ atoms cm$^{-3}$. The atoms have random motion in all the directions and their velocities are governed by Maxwell-Boltzmann distribution. The oven has an outlet in one of its walls so that the atoms moving towards this emerge out and move along the direction perpendicular to the orifice of the hole. The root mean square velocity of these atoms is given by

$$v_{rms} = \sqrt{3kT/m}. \tag{14}$$

For $T = 600$ K, $v_{rms} = \sqrt{2.48 \times 10^{-20}/m}$, which for $^{23}$Na turns out to be $\approx 800$ ms$^{-1}$. A small amount of the vapours is transferred into the highly evacuated glass cell where the remaining operations, including two stages of cooling, are carried out.

## 4.2. Laser cooling arrangement

The specially made cell has arrangement for six oppositely travelling laser beams oriented along the cartesian axes and intersecting at the middle of the cell. These laser beams are used to slow down the atoms to such an extent that their $v_{rms}$ is characteristic of temperature much lower than 1 K. Thus, for $T = 10^{-4}$ K or 100 μK, $v_{rms} = \sqrt{4.14 \times 10^{-27}/m}$ and for $^{23}$Na, we have $v_{rms} \approx 33$ cms$^{-1}$.

Now, suppose that an atom is moving along the right direction with instantaneous velocity $v$ and is shined with a laser beam travelling in the opposite direction. Scattering of incident photons from the atom will give it a kick in the direction opposite to that of its motion and, thus, reduce its speed. This is called radiation pressure.

But we know that atoms have electronic energy states, and at a temperature of about 600 K (which corresponds to a thermal energy of 0.05 eV) most of these will be in the lowest energy atomic state or at most very few in the first excited state. If the energy $h\nu$ of the incident photon matches the energy difference of the occupied level and the first excited state, then it will be absorbed completely and transfer its linear momentum $h\nu/c$ to the atom to reduce its speed. Of course, the atom cannot stay long in an excited state and will come back to the original ground state by spontaneous emission of a photon in a random direction and, thus, will have a random recoil momentum. This is the case of resonance scattering. Such multiple scatterings of about few millions per second, involve random recoils which cancel each other's effect to a great extent and lead to a net slowing down of the atoms. The main trick is that an atom absorbs photons travelling opposite to its own direction of motion. Note that this arrangement is akin to passing the atoms through a sticky medium and, hence, sometimes it is referred to as *Optical molasses*.

Having seen the importance of resonant scattering in cooling the gas down to much less than 1 K, we recall that moving atoms will have their eigen-energy separations shifted due to Doppler effect. Accordingly, the laser photon is resonantly absorbed if its energy is precisely the same as Doppler shifted energy of the first excited level; otherwise, it is just scattered transferring a small fraction of its energy and hence momentum. As the atoms are slowed down, the photon energy or frequency must be continuously tuned to match the changed Doppler shifted energy separation. The simplest way to achieve this purpose is to use a semiconductor diode laser. It is called *Chirp cooling*. Alternatively, the laser light frequency is kept fixed and the atomic beam is passed through a space varying but weak magnetic field so that the atomic energy separation gets changed with distance due to position dependent Zeeman effect. The magnetic field is continuously adjusted so that the Zeeman effect cancels the Doppler shift. The corresponding arrangement is known as *Zeeman slower*.

It may be mentioned that the minimum r.m.s. speed and hence the temperature attainable with laser cooling arrangement is limited by the fact that minimum momentum of the atoms will be of the order of the momentum carried by the photons, viz. $h\nu/c$. This is the so-called *recoil limit*. Consequently, the gas can be cooled at most down to a few μK or so.

The fact that six laser beams converge at the centre of the cell, photons from each of these push the absorbing atoms towards this point. In other words, the radiation pressure checks the tendency of the atoms to move away from the central part. Consequently, the gas gets confined in a region at the middle of the cell and atoms stay away from the walls besides getting cooled.

### 4.3. Magnetic trap and evaporative cooling

The temperature reached by laser cooling is not enough to produce BEC and the gas must be cooled further by few orders of magnitude. The first step in this direction is to turn off the lasers and use the inhomogeneous magnetic field based magnetic trap or magnetic bottle to proceed to achieve the goal.

It is well known that an atom with an unbalanced spin, as is the case with alkali atoms, has a magnetic moment $\boldsymbol{\mu}$. In the presence of an inhomogeneous magnetic flux density $\boldsymbol{B}$ with rate of change along z direction as $\frac{d\boldsymbol{B}}{dz}$, it experiences a force $\mathbf{F}$

$$\mathbf{F} = \hat{z}\left(\boldsymbol{\mu} \cdot \frac{d\boldsymbol{B}}{dz}\right), \tag{15}$$

where $\hat{z}$ is unit vector along z-direction. Now, the neutral atoms can be trapped only if the applied magnetic field has a local minimum and the atoms are low-field seekers, which is so if these have negative magnetic moment. The depth of the trap (in terms of energy) is given by $\boldsymbol{\mu} \cdot \boldsymbol{B}$ or $\mu B$. The magnetic moments of alkali atoms are negative and of the order of a Bohr magneton, $\mu_B = 9.27 \times 10^{-24}$ JT$^{-1}$ so that for a field of about 0.1 T, the trap depth is $\approx 10^{-24}$ J. This corresponds to a temperature $\approx \frac{10^{-24}}{1.38 \times 10^{-23}}$ K $\approx 0.07$ K. Thus, a gas of alkali atoms can be confined or stored in a magnetic trap using a field of 0.1T if its temperature is about 0.07 K. This explains why the gas must be precooled to few mK or less for being trapped magnetically.

Rather than going into the technical details of any of the different arrangements devised for trapping of atomic gases and making sure that losses of the gas particles are minimum, we would like to mention that the energy of the confined particles can be, in general, approximated by a nonhomogeneous harmonic potential so that the trap may be taken as an anisotropic magnetic bowl. The frequencies are determined by the applied magnetic field. Also, low-velocity atoms are close to the bottom of the potential and the atoms having high velocity are close to its top. The net effect of this trap is to confine the gas to a smaller region away from the walls of the cell and thus to compress it.

The gas stored in the trap is further cooled down to the desired ultra-low temperatures using forced evaporative cooling, which makes use of the fact that if the particles escaping from a system have an energy higher than their average energy, then the particles left behind re-thermalize to a lower average energy leading to cooling of the system.

One method of achieving this is to continuously decrease the trap depth by reducing the magnetic field intensity with time. Alternatively, the magnetic trap potential is kept constant and a radio-frequency (rf) field is used to continuously remove the higher energy atoms by precisely matching its frequency with the energy difference between the spin-up and spin-down states of the atoms in the magnetic field **B** of the trap. When rf field is switched on, the magnetic moment of the atoms near the top gets flipped. This makes the flipping atoms to be more energetic and enables these to escape the trap. The atoms left behind in the trap relax to have a distribution with lower total energy and hence characteristic of lower temperature. The intensity and frequency of the rf field are adjusted to be effective on the collection at lower temperature. This further reduces the temperature. A repetition of the process with properly reduced rf field frequency (every time) continually decreases the temperature of the atomic gas. As the atomic cloud cools, it shrinks towards the centre as velocity spread is reduced. This process cools the atomic gas to about $10^{-7}$ K or lower in about 20 – 30 seconds. The BE condensate is formed at the critical temperature of the system and the number of atoms in it increases with further reduction in temperature.

### 4.4. Recording the result

The linear dimensions of the BE condensate are generally very small. One way of observing this is to use a very powerful microscope and record the peak with the help of a camera attached with this using a flash of laser and phase contrast imaging technique. The shadow so obtained is then processed with a computer to get the final result.

The preferred method of detecting the condensate consists of measuring the momentum distribution of the atoms using time-of-flight approach. For this purpose, the confining magnetic field in the trap is suddenly turned off so that the cloud of ultracold atoms expands in the force free space as per their momentum or velocity distribution. Since at the prevailing temperatures of about 100 nK, the speed of the atoms is a few mms$^{-1}$ (see Eq. (14)), the cloud expands to about a few hundred mm in 100 μs or so. Note that the atoms move a distance proportional to their velocity so that the original shape of the peak is not changed. This is then shined with a laser pulse on resonance and recorded with the help of a CCD providing measure of the initial momentum distribution of the atoms.

The images recorded for different temperatures are interpreted to get the results. For temperatures above the critical temperature, the images are characteristic of isotropic distribution of velocities corresponding to thermal equilibrium; this corresponds to population of different energy states of the harmonic oscillator potential. On the other hand, for $T < T_c$, the shape of the condensate-image appearing at the middle is determined by the extent of non-homogeneousness of the trap field. The peak can be nearly round with a diameter of 10-50 μm, or a tapered cylinder with a diameter 10-20 μm and length few hundred microns. Further reduction in temperature makes the peak more intense. The atoms in the peak do not follow any thermal distribution indicating immensely close packing of atoms in the lowest energy state or the condensate, while the remaining atoms correspond to those populating the excited states with higher momenta and are governed by thermal distribution. Two computer-generated replicas of the image formed at two temperatures less than the critical temperature are depicted in Fig.2. At a lower temperature, the peak intensity increases at the cost of thermal cloud representing atoms in the excited states.

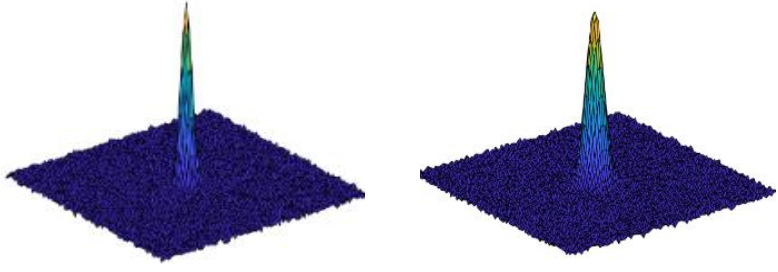

**Fig. 2.** Computer-generated replicas of BE condensate at two temperatures lower than the critical temperature; the image on the right depicts the situation at relatively lower temperature.

The peak is not infinitesimally narrow because of the uncertainty principle and the fact that the shape of condensate is controlled by the anisotropy of the trap magnetic field. It is pertinent to mention that the BEC peak is a remarkable example of the uncertainty principle being operative at macroscopic level.

An important consequence of anisotropy of the atomic gas in the trap is that BEC peak is observed not only in the momentum space but also in the coordinate space. This feature enlarges scope of experimental as well as theoretical investigations in BE condensates. It may be pointed out that this peculiarity is in contrast with the He I to He II transition and, also, the case of a uniform boson gas in a box, where condensation occurs only in the momentum space and the particles in the ground as well as the excited states fill the same volume. Indeed, condensate peaks over a broad distribution have been recorded in both the velocity and space distributions.

## 5. THEORETICAL CONSIDERATIONS FOR ULTRACOLD ALKALI ATOMS

It has been pointed out in the preceding section that the alkali atoms for BEC are stored in a magnetic trap rather than a box of volume V. These atoms experience an inhomogeneous potential, which can be written as

$$V(\mathbf{r}) = \frac{m}{2}(\omega_x^2 x^2 + \omega_y^2 y^2 + \omega_z^2 z^2), \qquad (16)$$

where the angular frequencies $\omega_l$ are determined by the trap magnetic field. The energy eigenvalues are given by

$$\varepsilon_{i_x i_y i_z} = \left(i_x + \frac{1}{2}\right)\hbar\omega_x + \left(i_y + \frac{1}{2}\right)\hbar\omega_y + \left(i_z + \frac{1}{2}\right)\hbar\omega_z, \qquad (17)$$

with $i_l = 0,1,2,\ldots$ as the quantum numbers. Using $\varepsilon_{i_x i_y i_z}$ in place of $\varepsilon_j$ in Eq. (2), we get expression for the mean occupation number for the relevant oscillator energy state. Assuming $N$ to be so large that $i_l$ can have quite large values, zero-point energy in Eq. (17) becomes irrelevant. Then it is found that an ideal gas in such a potential will undergo BEC at critical temperature [8-10]

$$T_c^{ho} = \frac{0.941\, \hbar\omega_{ho}}{k} N^{1/3} = 7.2 \times 10^{-12} \omega_{ho} N^{1/3}, \tag{18}$$

where $\omega_{ho} = (\omega_x \omega_y \omega_z)^{1/3}$ is the geometrical mean of the oscillator frequencies along the three cartesian axes. Furthermore, the number of bosons in the lowest energy state $\varepsilon_{000}$, is given by

$$N_0^{ho} = N - 1.202 \left(\frac{kT}{\hbar\omega_{ho}}\right)^3. \tag{19}$$

Combining Eqs. (18) and (19), we get the temperature dependence of the condensate fraction for $T \leq T_c^{ho}$ as

$$\frac{N_0^{ho}}{N} = 1 - \left(\frac{T}{T_c^{ho}}\right)^3. \tag{20}$$

The difference between these three expressions for the bosons in the inhomogeneous field and the corresponding relations for the uniform gas in a box of volume $V$ is quite clear. However, the physical arguments presented in Section 2 still hold good. It may be mentioned that in this case, the thermodynamic limit corresponds to the situation $N \to \infty, \omega_{ho} \to 0$ while the product $N\omega_{ho}^3$ is finite. In the first experimental observation in $^{87}$Rb gas, $\nu_{ho}$ was 120 Hz and $N$ was 2 x $10^4$, for which Eq. (18) yields $T_c^{ho}$ = 147 nK, which is reasonably close to the experimental value of nearly 170 nK. It is important to note that the condensate consisted of the coldest atoms in the world.

It is worth mentioning that the above simple model for study of BEC in alkali atomic gas has been modified to accommodate the effect of (i) finite size of the system, and thus its departure from the thermodynamic limit, as $N$ values range from $10^3$ to $10^9$ and $\omega_{ho}$ is few hundred; and (ii) interaction between the gas atoms. The size-correction, which also includes the contribution of zero-point energy, reduces both the transition temperature and the condensate fraction. The decrease in $T_c^{ho}$ is by a factor that depends on the values of frequencies and varies with N as $N^{-1/3}$, while the reduction in $\frac{N_0^{ho}}{N}$ is determined by these two factors and $\left(\frac{T}{T_c^{ho}}\right)^2$. Thus, the corrections due to finite number of gas atoms are small but observable (few percent). It has also been established that though the gas has quite low number density, the interactions have reasonable effect on their various properties. As a first approximation, the effect of the interactions in the alkali atoms condensate, is considered through the Gross-Pitaevskii equation, which is essentially nonlinear Schrodinger equation. Some other approaches have also been followed. The main finding is that inclusion of interaction effects further diminishes the values of $T_c^{ho}$ and $\frac{N_0^{ho}}{N}$. On the whole, the two modifications significantly improve agreement between the experimental results for condensate fraction and the theory as compared to the thermodynamic result [10].

Before proceeding further, we have a look at the variation of condensate fraction with the scaled temperature. The plots obtained from Eqs. (12) and (20) have been depicted in Fig. 3. Also included herein are some points extracted from the figure for experimental results reported by Ensher et al [11]. They had found that the least square fit of their experimental data for about 25 values of $\frac{T}{T_c^{ho}}$ less than unity, to theoretical expression, Eq. (20), leads to a modified critical temperature which is 0.94 times $T_c^{ho}$, implying a total correction due to finite size and interactions to $T_c^{ho}$ value to be about 6%. The corresponding curve has also been included as black solid line in Fig. 3.

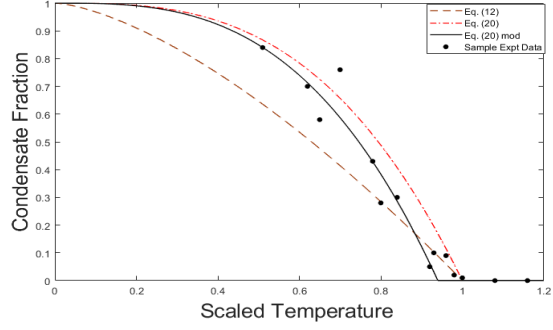

**Fig. 3** Plots showing variation of condensate fraction with scaled temperature as given by Eq. (12), Eq. (20) and modified version of Eq. (20). Also included are some sample experimental points extracted from the relevant figure in [11].

## 5. Epilogue

The discovery of BEC triggered a fervour for observing this phenomenon in different types of bosons and for investigating its properties. Besides spin polarized hydrogen and alkali atoms, it has been obtained in some alkaline earth and lanthanide atoms, few molecules, quasiparticles, etc. The multitude of isotopes making this list are $^1$H, $^7$Li, $^{23}$Na, $^{39}$K, $^{41}$K, $^{85}$Rb, $^{87}$Rb, $^{133}$Cs, $^{40}$Ca, $^{52}$Cr, $^{84}$Sr, $^{86}$Sr, $^{88}$Sr, $^{164}$Dy, $^{168}$Er, $^{170}$Yb and $^{174}$Yb. Besides, BEC has been reported to form in the metastable (excited) state of $^4$He gas. Note that $^{52}$Cr is the first element in group VI B. Some of the diatomic molecules of otherwise fermionic atoms, which have been found to undergo BEC transition include $^6$Li$_2$, $^{40}$K$_2$. Very recently, BEC has been obtained in $^{87}$Rb atoms in space under low gravity conditions and results of different experiments have been compared with those on the ground. One important conclusion is that the fraction $\frac{N_0^{ho}}{N}$ in space has substantially higher value than that at the ground [12].

The presence of mass m in the denominator of the expression for $T_c$, Eq. (10), implied that it would be possible to obtain a condensate at elevated temperatures, as high as the room temperature, in the systems with lower mass. This made researchers to look for BEC in various integer-spin quasiparticles in solids because their effective mass is even less than that of an electron. Indeed, this has been observed in magnons, excitons, exciton-polaritons, surface plasmon polaritons, and optical phonons using different novel techniques which generally increase the boson concentration above a critical value. These pursuits are stimulated by the hope that condensates obtained at temperatures close to the room temperature will help in developing BEC based technology for commercial purpose. One such encouraging example is realization of BEC in the magnons in thin film of ferromagnetic yttrium-iron-garnet at room temperature. However, till date the quantity of the condensate obtained is so low that going beyond laboratory appears to be a dream.

Interestingly, BEC has been reported even in photons using a fluorescent dye-filled optical microcavity as well as few meters long erbium-ytterbium co-doped fibre cavity [13]. These arrangements effectively act as white-wall photon containers, which is in complete contrast with a black body for which photon number is not conserved, which makes $\mu = 0$.

Recall that a BE condensate can be described by a single wavefunction corresponding to complete in-phase superposition of the wavefunctions or de Broglie waves of the constituent bosons in the lowest energy state. Thus, the condensate consists of $N_0$ coherent entities. Accordingly, like an optical laser which gives coherent photons, an atomic or molecular gas condensate constitutes a coherent source of matter waves and is, therefore, referred to as atomic or molecular laser. The trap is analogous to cavity in the conventional laser. However, the two differ in the sense that a matter wave laser is in thermal equilibrium at very low temperature, while the optical laser operates in nonequilibrium condition characterized by population inversion. Nonetheless, interference patterns, consisting of alternate stripes of high and low number density, formed by superposition of two or more parts of condensates obtained from a trap were reported in 1997. In addition, it was shown that a part of the condensate could be extracted to form a sort of pulsed atom laser. Also, all the nonlinear optics effects observed with photons can be obtained with matter wave lasers. Obviously, an atomic / molecular laser provides an interferometer-based arrangement for a variety of precise measurements through small fringe shifts – the quantum metrology. It is expected that the studies related to condensates will go a long way in quantum information processing and development of quantum computers.

In addition to showing interference, the condensate has been found to give Bragg diffraction from a lattice formed by two overlapping laser beams. Furthermore, one can generate quantized vortices as well as matter wave solitons in the condensates. Some experiments involving BEC have also led to new insights in many areas of condensed matter physics.

Recently, Eckel and colleagues [14] have reported results of experimental and theoretical studies on the dynamics of a supersonically expanding, ring shaped BE condensate. They have found these to be quite comparable to the expanding universe, and, thus, have brought out a possible use of the condensates in the study of cosmological phenomena.

We conclude the article by quoting Snoke and Baym, who, in the Introduction to book 'Bose-Einstein Condensation' edited by Griffin, Snoke and Stringari (Cambridge University Press,1995), wrote, 'in its broadest sense BEC is a common phenomenon occurring in physics on all scales, from condensed matter to nuclear, elementary particles, and astrophysics, with ideas flowing across boundaries between fields'.